\documentclass[aps,prl,preprint,superscriptaddress
]{revtex4-2}

\usepackage{multirow}
\usepackage{amsmath,amssymb,amsfonts}
\usepackage{amsthm}
\usepackage{mathrsfs}
\usepackage[title]{appendix}
\usepackage{xcolor}
\usepackage{textcomp}

\usepackage{booktabs}
\usepackage[utf8]{inputenc}
\usepackage{algorithm}
\usepackage{algorithmicx}
\usepackage{algpseudocode}
\usepackage{listings}
\usepackage{caption}
\usepackage{subcaption}
\usepackage{graphicx}
\usepackage{dcolumn}
\usepackage{bm}

\begin{document}

\preprint{}

\title{Characterizing Neon Thin Film Growth with an NbTiN Superconducting Resonator Array}

\author{Kyle Matkovic}
\affiliation{
School of Physics, UNSW Sydney, NSW 2052, Australia
}%
\author{Patrick Russell}
\affiliation{
School of Electrical Engineering, UNSW Sydney, NSW 2052, Australia
}%
\author{Andrew Palmer}
\affiliation{
School of Physics, UNSW Sydney, NSW 2052, Australia
}%
\author{Eric Helgemo}
\affiliation{
School of Physics, UNSW Sydney, NSW 2052, Australia
}%
\author{Lukas Delventhal}
\affiliation{
School of Physics, UNSW Sydney, NSW 2052, Australia
}%
\author{Kun Zuo}
\affiliation{
School of Physics, University of Sydney, NSW 2006, Australia
}%
\author{Kundan Surse}
\affiliation{
School of Physics, UNSW Sydney, NSW 2052, Australia
}%

\author{Rajib Rahman}
\affiliation{
School of Physics, UNSW Sydney, NSW 2052, Australia
}%

\author{Maja C. Cassidy}
\affiliation{
School of Physics, UNSW Sydney, NSW 2052, Australia
}%
\email{maja.cassidy@unsw.edu.au}

\date{\today}

\begin{abstract}
Electrons levitating above the surface of solid neon have recently emerged as a promising platform for high-quality qubits. The morphology and uniformity of the neon growth in these systems is crucial for qubit performance in a scalable architecture.  Here we report on the controlled growth and characterization of thin solid neon films using multiplexed superconducting microwave resonators.  By monitoring changes in the resonant frequency and internal quality factor ($Q_i$) of an array of frequency multiplexed quarter-wave coplanar waveguide resonators, we quantify the spatial uniformity of the film. A pulsed gas deposition protocol near the neon triple point results in repeatable film formation, generating measurable shifts in frequency and variations in $Q_i$ across the resonator array. Notably, introducing a post-deposition anneal at 12 K for one hour improves the film homogeneity, as shown by the reduced resonator-to-resonator variance in frequency and $Q_i$, consistent with enhanced wetting. These results demonstrate resonator-based metrology as an in-situ tool for characterising neon film growth, directly supporting the development of electron on inert quantum solid qubit platforms.  
\end{abstract}
\maketitle


Electrons levitating on condensed noble gases present a promising platform for hosting qubits owing to the ultraclean environment that suppresses common mechanisms for decoherence \cite{Platzman99}. A variety of qubit implementations have been explored, utilizing the spin \cite{Lyon06}, charge \cite{Schuster10}, or Rydberg state \cite{Kawakami24} of the electron. This levitation arises from the interplay between attractive image potentials and repulsive Pauli exclusion forces, which confine the electron at a stable distance above the surface \cite{Cole69}. Among the proposed implementations, solid neon stands out as an optimal substrate due to its weak electron-surface interaction and the stability of its solid phase \cite{zavyalov_electron_2005}. Recent breakthroughs have demonstrated single electron charge qubits levitating on solid neon \cite{Zhou22}, achieving coherence times exceeding 0.1 ms alongside fast nanosecond-scale gate operations \cite{Zhou24} and multi-qubit operation \cite{Li25_2}. Efforts are now focused on scaling these qubits and enabling their operations at higher temperatures \cite{Li25}.

A key requirement for further progress is the reliable fabrication of high-quality solid neon films on patterned substrates. However, there remains a lack of in situ characterizing and diagnostic tools compatible with the hermetic sample cell at cryogenic temperatures. While bulk neon crystallizes into a face centered cubic (fcc) lattice below 24.54 K \cite{Hill11}, with large single crystals achievable\cite{Endoh75}, thin neon films behave quite differently. Optical quartz microbalance measurements reveal that the growth of such films differs dramatically from the bulk behaviour\cite{Pandit83}. Crystalline neon films grown on surfaces by lowering the temperature through the triple point are only thermodynamically stable for a few monolayers, and the film thickness diverges as the saturated vapor pressure is approached, a phenomenon known as triple point wetting \cite{Migone86}\cite{Leiderer92}. Achieving continuous films on structured surfaces, as is required for qubit applications, therefore presents a significant challenge. Alternative approaches, such as quench condensation, can produce thicker films but often introduce a higher density of crystal defects~\cite{Leiderer92}. For quantum computing architectures that require physically moving electrons, such as spin shuttles \cite{taylor_fault-tolerant_2005}, or implementation of long range error correcting codes \cite{doi:10.1126/sciadv.abn1717}, even minor inhomogeneity or defects in the film could be sources of decoherence and thus detrimental. Achieving uniform, low-defect neon films on structured substrates is therefore essential for continued development and scaling of these qubits.

In this work, we utilize high quality factor superconducting resonators as a diagnostic tool to probe the in-situ uniformity of the neon film deposition. These resonators are extremely sensitive to local environmental changes, and by frequency multiplexing several resonators to a single transmission line, we can construct a spatial map of the entire chip area.  A change in the resonator capacitance causes a shift in both the resonant frequency and internal quality factor, which allows for an estimate of the film thickness and defect density across the resonator area $\sim1\ \text{mm}^2$.  We find that introducing a high temperature anneal at $\sim$12 K for 1 hour significantly improves the uniformity of neon film growth, reducing the variance in relative frequency shifts by three orders of magnitude.

Figure 1a shows a rendering of the measured device. The design consists of eight quarter-wave superconducting coplanar waveguide (CPW) resonators multiplexed to a single transmission line. Resonators 1-4 in the top row had central conductor width w = 6 $\mu$m, and separation to the groundplane d = 5 $\mu$m,  while resonators 5-8 in the bottom row had w = 10 $\mu$m and d = 8 $\mu$m, resulting in a designed impedance of $\sim$55 $\Omega$ ($Z = \sqrt{L/C}$, and resonant frequency 
$f_r = \frac{1}{2\pi\sqrt{LC}}$, where $L$ is the resonator inductance and $C$ is the resonator capacitance.  Capacitive coupling between the resonators and feedline was achieved by bringing a segment of 200 $\mu$m of the resonator in proximity to the feedline, separated by 35$\mu$m of ground plane. Using an analytical model from  \cite{Besedin18}, we extract a designed coupling capacitance of $C_c \sim 0.6$ fF.
The sample was fabricated from a film of 40nm NbTiN reactively sputtered from a high purity Nb-Ti target in an Ar-N atmosphere onto a high resistivity silicon wafer (Topsil). The resonators were defined using electron beam lithography before being dry etched in a $SF_6/O_2$ environment followed by a solvent clean.  After dicing, the sample was wire bonded to a printed circuit board and mounted in a light-tight, hermetically sealed sample cell attached to the mixing chamber of a dilution refrigerator similar to \cite{Zhou22}. 

The microwave transmission through the device was measured with a vector network analyzer with a standard circuit QED setup within a dilution refrigerator (BlueFors LD) with a base temperature of 20mK. The resonators were first fully characterized at mK temperatures before the sample was warmed to 26 K for neon deposition, close to the boiling point of 27.104 K. Neon (5N purity gas) was delivered to the sample by a stainless steel capillary line (internal diameter 0.18 mm, VICI JOUR) connected to the cell. To deposit neon on chip, a small volume between two solenoid valves was pressurized to ~11 Bar and opened to the capillary line to control the amount of gas entering the sample cell. The cell was held just above the neon triple point of 24.54 K, with the 4K plate at $\sim 27$ K, the still plate at $\sim 26$ K and the mixing chamber and sample at 26 K, so that the fixed quantity of gas was deposited as a liquid layer on the chip surface. After a delay of two seconds for the gas to liquefy, the volume was closed off from the capillary line and refilled, repeating the process for 50 pulses providing a consistent neon deposition. After deposition, the sample was cooled to 4K, where the temperature was held for ~10 hours, and then the fridge condensed to 20 mK for measurement. For some measurements, the cool down was stopped at an intermediate temperature of $12~K$ allowing for an additional anneal to take place.

\begin{figure}[ht]
    \centering
    \includegraphics[width=0.8\textwidth]{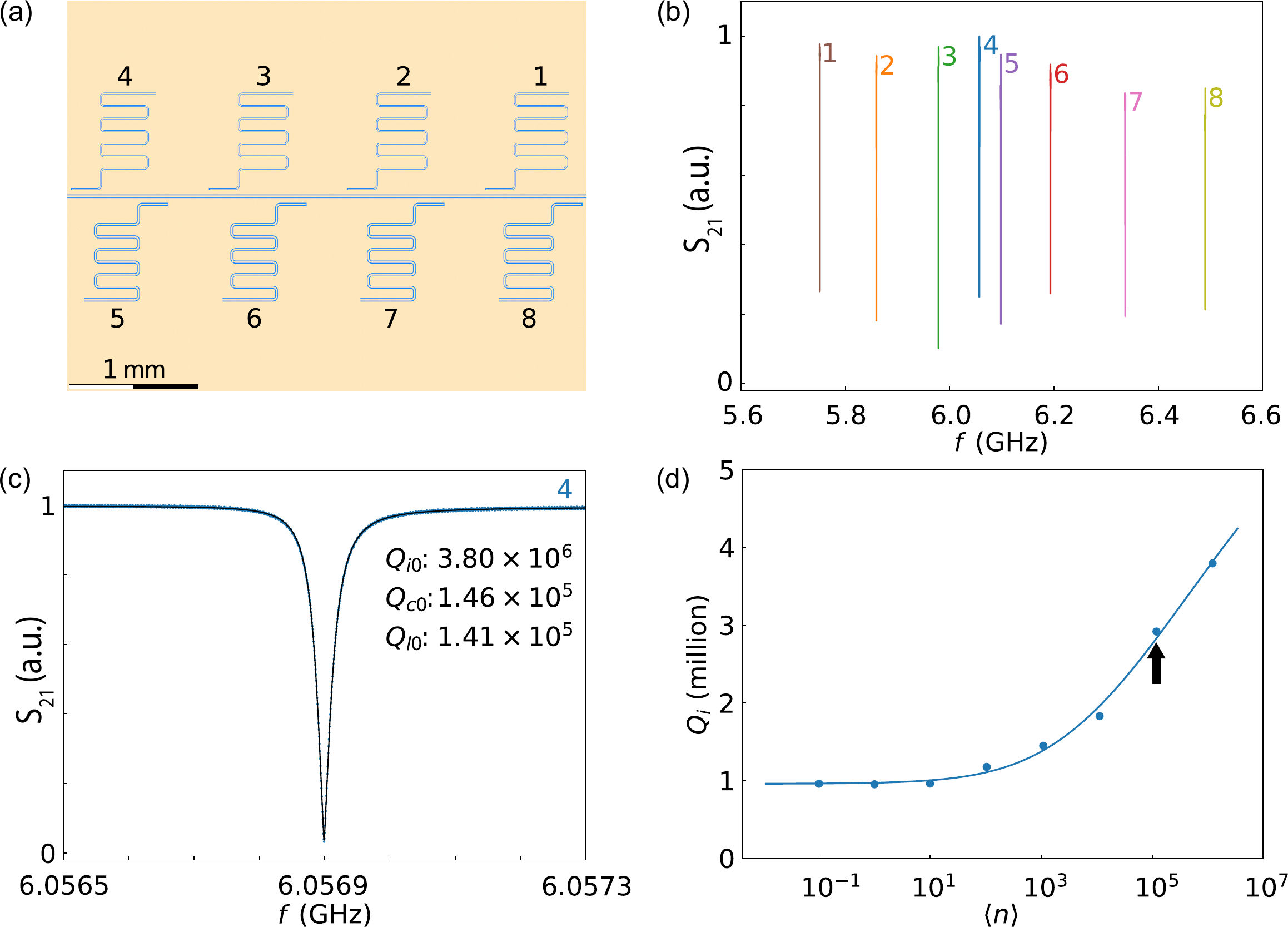}
    \label{fig:1}
    \caption{(a) Render of the multiplexed hanger resonator device and (b) microwave transmission measured as a function of frequency, showing 8 distinct resonances. (c) Zoom in of resonator 4, with fit as described in Eq. \ref{eqn:s21}. (d) Extracted internal quality factor for resonator 4 before the second deposition with anneal, as a function of photon number, together with a fit to the data given by Eq. \ref{eqn:qi}. The regime studied in this paper is indicated by the black arrow. }
\end{figure}
The microwave transmission spectrum ($S_{21}$) shows eight distinct resonances with frequencies between 5.7-6.5 GHz, as shown in Fig 1 b. Zooming in on one of these resonances allows for the transmission spectra to be fit \cite{Probst15} by the equation

\begin{equation}\label{eqn:s21}
S_{21} = A e^{i \alpha} e^{-2 \pi i f \tau}\left(1-\frac{\frac{Q_l}{|Q_e|}e^{i\phi}}{1+2iQ_l\frac{f-f_r}{f_r}}\right).
\end{equation}

Here $A$ is the amplitude of the bare feedline transmission in the absence of a resonance, $\alpha$ is a background phase shift parameter to account for non-uniform background transmission, $\tau$ is the electronic delay caused by the non-zero time it takes light to travel through the cables, $f$ is the probe frequency, $f_r$ is the resonance frequency, $Q_l$ is the loaded quality factor, $Q_e$ is a complex-valued external quality factor related to $Q_c$ via $1/Q_c = \text{Re}(1/Q_e)$, and $\phi$ captures any impedance mismatch between the resonator and feedline. Fitting this model to the S21 data, we are can identify the internal quality factor $Q_i$ from
\begin{equation}
    \frac{1}{Q_i} = \frac{1}{Q_l} - \frac{1}{Q_c}.
\end{equation}
 Measured internal quality factors ($\langle n \rangle \approx 10^5$) for the resonators range from $4.56 \times 10^5$ to $1.49 \times 10^6$ with coupling $Q_c$ between $8.0 \times 10^4$ and $3.0 \times 10^5$ as displayed in Tab. \ref{tab}. Varying the microwave power at the resonator shows an increase in internal quality factor at high power due to saturation of two level systems. The extracted $Q_i$ as a function of photon occupancy of the resonator is then fit the model \cite{Crowley23}
\begin{equation}\label{eqn:qi}
    \frac{1}{Q_i} = \frac{1}{Q_\text{TLS}(\langle n \rangle)} + \frac{1}{Q_\text{oth}},  
\end{equation}
where $Q_\text{TLS}$ is the quality factor associated with two-level system (TLS) losses which depends on $\langle n \rangle$ the average photon number supplied to the resonator and $Q_\text{oth}$ considers any other power independent loss sources. The results presented in this paper came from studying the resonators in the high power regime at $10^5$ photons, as indicated by the arrow in Figure 1d.

Figure 2a shows a measurement of resonator 4 before and after neon deposition. Post deposition, the resonance displays a shift towards lower frequencies due to the increased capacitance from the deposited neon. A deepening of the resonance is also observed, which corresponds to an improvement in the internal quality factor $Q_i$. From this, we estimate a neon thickness of 28 nm on the surface of the resonator. Figure 2b presents the change in quality factor $\Delta Q_i$ and the relative changes in frequency $\Delta f_r / f_{r0}$, where $\Delta f_r$ is the change in frequency after neon deposition. We observe quality factor improvements across all resonators after neon deposition up to 135 \% compared to before neon deposition (see Table 1) and that the strength of this affect appears related to the strength of the frequency shift.

\begin{figure}[t]
    \centering
    \includegraphics[width=0.8\textwidth]{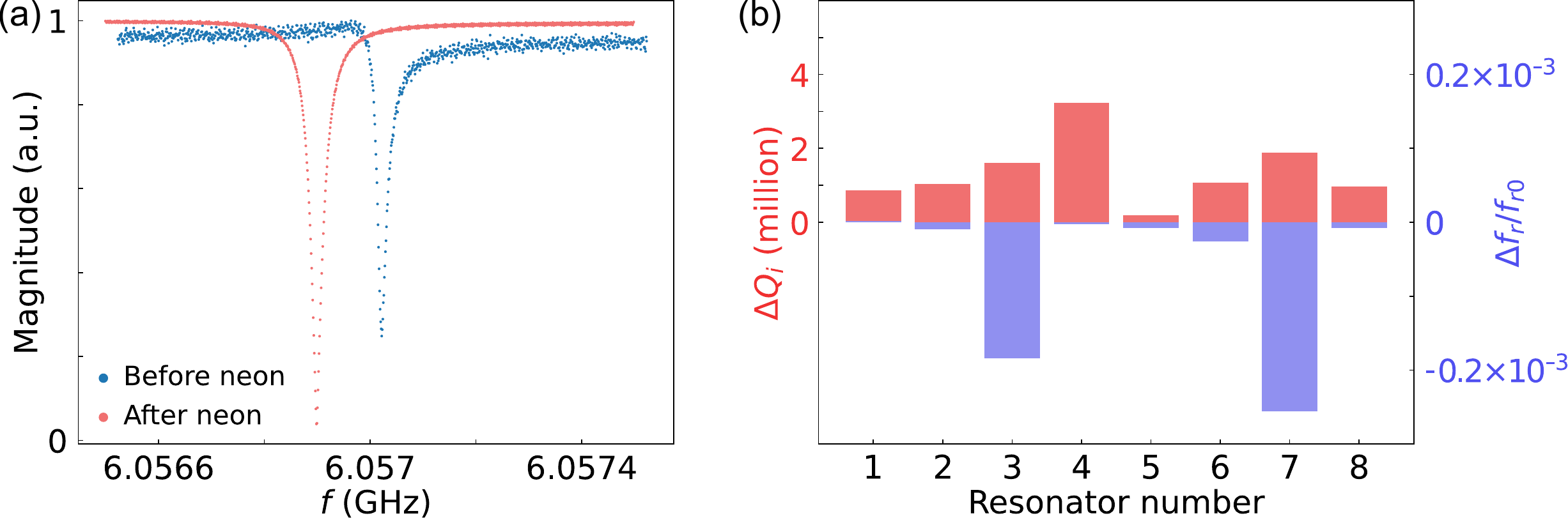}
    \label{fig:shifts}
    \caption{(a) $S_{21}$ measurement of resonator 4 before and after neon deposition (without anneal). (b) Relative shift in frequency and absolute change in $Q_i$ before and after neon deposition for each resonator. }
\end{figure}

\begin{figure}[ht]
    \centering
    \includegraphics[width=0.8\textwidth]{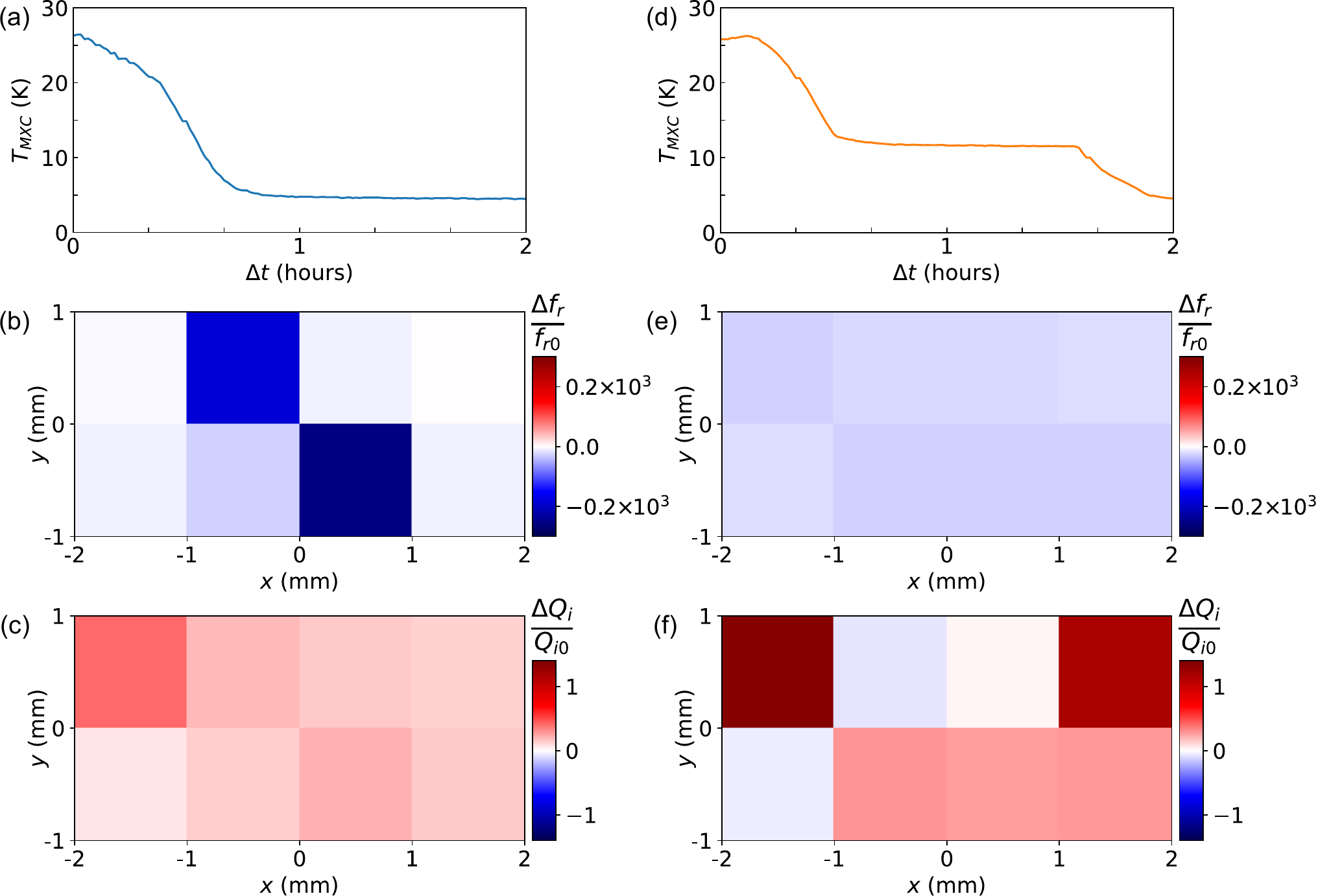}
    \label{fig:anneal}
    \caption{Relative changes in frequency and internal quality factor mapped to the resonator locations on chip for cooldowns without, (a-c), and with, (d-f), a high temperature anneal at 12~K. (a) and (d) display the mixing chamber plate temperature as a function of time during the cooldowns. (b) and (e) show the relative changes in frequency, and (c) and (f) show the relative changes in $Q_i$ post neon deposition, mapped onto the location on the chip. }
\end{figure}

\begin{table}[h]

\caption{Changes in frequency and $Q_i$ for cooldowns without and with anneal for resonators 1 to 8. Relative changes in frequency are defined as the ratio over the change in frequency $\Delta f_r$ over the frequency before deposition $f_{r0}$. }\label{tab}
\begin{tabular*}{\textwidth}{@{\extracolsep\fill}lcccccccccc}
\toprule%
& \multicolumn{4}{@{}c@{}}{Initial values} & \multicolumn{3}{@{}c@{}}{Without anneal} & \multicolumn{3}{@{}c@{}}{With anneal} \\\cmidrule{2-5}\cmidrule{6-8}\cmidrule{9-11}%
 & $f_{r0}$ \footnotesize /GHz & $Q_{i0}$ & $Q_{c0}$ & $Q_{l0}$ & $\frac{\Delta f_r}{f_{r0}}$ &$t_{Ne} [nm]$& $\Delta Q_i$ &  $\frac{\Delta f_r}{f_{r0}}$ &$t_{Ne} [nm]$& $\Delta Q_i$ \\
\midrule
1  & 5.75126 & 7.04 E+5 & 2.8 E+5 & 2.0 E+5 & 1.56 E-6  & -   & 8.65 E+4 & -2.10 E-5 & 25 & 4.43 E+6\\
2  & 5.85966 & 7.12 E+5 & 1.8 E+5 & 1.4 E+5 & -9.30 E-6 & 12  & 1.03 E+5 & -2.31 E-5 & 27 &4.13 E+4 \\
3  & 5.97875 & 1.49 E+6 & 1.2 E+5 & 1.0 E+5 & -1.84 E-4 & 160 & 1.60 E+5 & -2.19 E-5 & 26 &-6.07 E+4 \\
4  & 6.05701 & 9.22 E+5 & 3.0 E+5 & 2.2 E+5 & -2.93 E-6 &  5  & 3.23 E+5 & -2.59 E-5 & 30 &5.17 E+6 \\
5  & 6.09828 & 4.56 E+5 & 8.0 E+4 & 6.3 E+4 & -7.29 E-6 &  10 & 1.98 E+4 & -2.06 E-5 & 25 &-1.09 E+4 \\
6  & 6.19305 & 8.76 E+5 & 3.7 E+5 & 2.5 E+5 & -2.61 E-5 & 30  & 1.07 E+5 & -2.81 E-5 & 32 &4.85 E+5 \\
7  & 6.33645 & 8.94 E+5 & 2.9 E+5 & 2.2 E+5 & -2.55 E-4 & 210 & 1.89 E+5 & -2.61 E-5 & 30 &1.37 E+5 \\
8  & 6.48971 & 1.01 E+6 & 2.9 E+5 & 2.0 E+5 & -7.88 E-6 & 11  & 9.69 E+4 & -2.63 E-5 & 30 &8.49 E+4 \\
\botrule
\end{tabular*}
\end{table}

Although the freezing temperature of bulk neon crystals is 25~K, atomic rearrangements can take place well below this temperature in polycrystalline or porous films where the binding energy of each atom to the lattice is less. Previous investigations of quench-condensed neon films on bulk substrates have shown a structural rearrangement in the atoms at intermediate temperatures in the range  1 - 4 K \cite{CLASSEN1999163}, with an increase in stiffness and decrease in the density of tunneling states observed at higher annealing temperatures. Complete desorption of quench-condensed neon films has been shown to occur at temperatures above 8.5~K  \cite{PhysRevLett.88.016104}, with the material then shown to recondense across the substrate. We chose an annealing temperature of $12~K$, above this desorption temperature but well below the melting point of bulk neon, to investigate the effects of annealing on the spatial dependence of the neon uniformity across our patterned resonator samples.

Figure 3 shows the relative frequency shift and internal quality factor improvement as a function of resonator position on the chip. For depositions without an additional high temperature anneal (Fig 3 a-c), the frequency response is quite different for each resonator, ranging from a relative change of $1.56 \times 10^{-6}$ to $-2.55 \times 10^{-4}$ as shown in Tab. 1. Larger changes in frequency are localised about the centre of the chip indicating incomplete wetting of the neon film across the surface. For depositions with an additional anneal at 12 K (Fig 3 d-f), the relative frequency changes across the chip are smaller (between $-2.06 \times 10^{-5}$ to $-2.81 \times 10^{-5}$), indicating a more uniform coverage of the surface. Suprisingly, we observe the strongest improvements in $Q_i$ occur where the frequency changes the least.  The Qi improvements persist even when the neon is removed in a subsequent cooldown, indicating that the neon `cleans' the surface of the chip by removing contaminants from the surface. Further understanding of this effect will be investigated in future work.

\begin{figure}[H]
  \centering

    \includegraphics[width=0.9\textwidth]{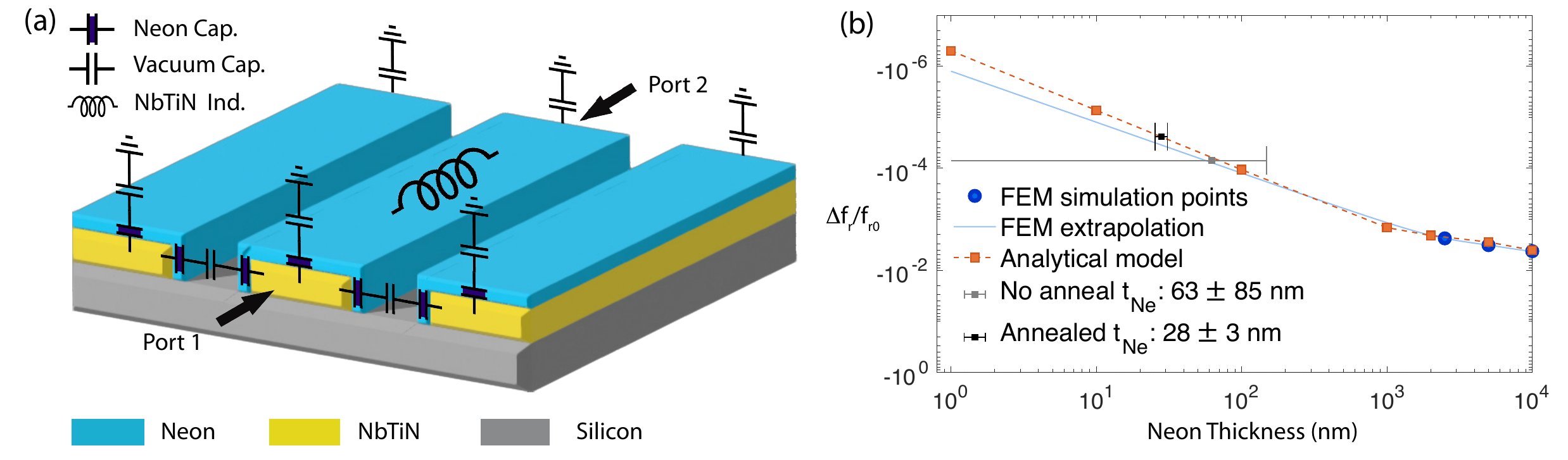}
    \caption{(a) Schematic of the analytical model of the neon-covered CPW with capacitors and inductors, (b) Estimation of Neon thickness from the analytical model and FEM simulation with Neon thickness before and after annealing }
  \label{fig:Simulation Plot}
\end{figure}

To determine the neon layer thickness, we adopt an analytical modelling approach to overcome the challenges presented capturing nanometer and millimeter-scale dimensions using finite element modeling (FEM) approaches. An equivalent circuit model, comprising inductors and capacitors, is developed to represent the coplanar waveguide, thereby allowing the investigation of arbitrarily small neon layer thicknesses. In this model, the distributed impedance of the coplanar waveguide is realised using five serially connected $\pi$–CLC networks, as illustrated in Figure \ref{fig:Simulation Plot} (a). The effect of the neon layer is incorporated by introducing an additional capacitance filled with neon, using its dielectric constant of 1.2445, which accurately captures the influence of the dielectric layer within the analytical framework. The circuit-level model is implemented and simulated in Keysight ADS software, where S-parameter analysis is performed to evaluate the response. To validate the analytical approach, FEM simulations were carried out in Keysight ADS for larger neon thicknesses of $2.5 \mu m, \ 5\mu m \ and \ 10 \mu m $. The extrapolated FEM results exhibit excellent agreement with the analytical predictions, as shown in Figure \ref{fig:Simulation Plot} (b). From this, we extract an estimated neon thickness ranging from 5 to 210 nm (mean 63 nm) across the resonator array without annealing, which was significantly reduced to a uniform range of 25–32 nm (mean 28 nm) following the annealing process. 

This work demonstrates that multiplexed superconducting resonators can serve as a highly sensitive tool for monitoring and optimizing thin neon film growth, a critical element for developing electron-on-neon qubit systems. The observed frequency shifts provide clear evidence that combining the neon deposition with an intermediate temperature anneal at 12 K leads to more homogeneous films with more complete wetting of the chip surface.  The observed improvements in internal quality suggest that neon films can help ``clean" the chip surface, enhancing resonator performance. While the size scale of these CPW resonators limit the resolution over which the neon growth can be studied to ~500 x 500 um areas, high-density lumped element resonators could reduce this by at least another order of magnitude \cite{PhysRevResearch.5.043126}. Making the resonators out of high $T_c$ superconductors, such as YBCO, would allow for the study of the neon deposition and uniformity through the neon deposition proccess. These results establish a robust and reproducible method of characterising neon film growth, providing a new tool to support the robustness and scalability of electron-on-neon qubit architectures.



\begin{acknowledgments}
We thank Dafei Jin, Xinhao Li, Erika Kawakami and Ivan Grytsenko for assistance in establishing the neon experiments in our lab, and Paul Leiderer and Chrisian Enss for helpful discussions. K. Matkovic and P. Russell acknowledge the support of Sydney Quantum Academy Undergraduate Summer Scholarships, and CSIRO Next-Gen Undergraduate Honours Scholarships. K. Zuo acknowledges support from the ARC Center of Excellence for Engineered Quantum Systems (CE170100009),   A. Palmer acknowledges support from the CSIRO Next-Gen Graduates Program,  L. Delventhal acknowledges support from the Sydney Quantum Academy and a UNSW Tuition Fee Scholarship, M. C. Cassidy acknowledges support from a UNSW Scientia Fellowship and an Australian Research Council Discovery Early Career Research Fellowship (DE240100590).  The authors acknowledge the facilities as well as the technical assistance of the Research and Prototype Foundry Core Research Facility at the University of Sydney, part of the NSW node of the NCRIS-enabled Australian National Fabrication Facility.
\end{acknowledgments}


\bibliography{fshift}

\end{document}